\documentclass{hotnets22}

\usepackage{times}  
\usepackage{hyperref}
\usepackage{titlesec}
\usepackage{algorithmic}
\usepackage{algorithm}
\usepackage{listings}
\usepackage{graphicx}
\usepackage{textcomp}
\usepackage{xcolor}
\usepackage{url}
\usepackage[capitalise]{cleveref}

\hypersetup{pdfstartview=FitH,pdfpagelayout=SinglePage}

\begin{document}

\title{bypass4netns: Accelerating TCP/IP Communications in Rootless Containers}

\numberofauthors{2}
\author{
\alignauthor
Naoki Matsumoto\\
       \affaddr{Kyoto University}\\
       \affaddr{Kyoto, Japan}\\
       \email{mt2.naoki@inet.media.kyoto-u.ac.jp}
\alignauthor
Akihiro Suda\\
       \affaddr{NTT Software Innovation Center}\\
       \affaddr{Tokyo, Japan}\\
       \email{akihiro.suda.cz@hco.ntt.co.jp}
}

\maketitle

\begin{abstract}
"Rootless containers" is a concept to run the entire container runtimes and containers without the root privileges.
It protects the host environment from attackers exploiting container runtime vulnerabilities.
However, when rootless containers communicate with external endpoints, 
the network performance is low compared to rootful containers because of the overhead of rootless networking components.

In this paper, we propose \textbf{bypass4netns} that accelerates TCP/IP communications in rootless containers by bypassing slow networking components.
bypass4netns uses sockets allocated on the host.
It switches sockets in containers to the host's sockets by intercepting syscalls and injecting the file descriptors using Seccomp.
Our method with Seccomp can handle statically linked applications that previous works could not handle.
Also, we propose high-performance rootless multi-node communication.
We confirmed that rootless containers with bypass4netns achieve more than 30x faster throughput than rootless containers without it.
In addition, we evaluated performance with applications and it showed large improvements on some applications.
\end{abstract}

\section{Introduction}
The container is an isolation technology provided by container runtimes with Linux kernel features.
They have the advantage of higher speed performance and more efficient use of computing resources compared to virtual machines.
A software stack that provides containers consists of low-level runtimes(e.g. runc\cite{runc}), high-level runtimes(e.g. containerd\cite{containerd}), and container engines(e.g. Docker\cite{docker} and Podman\cite{podman}).
Processes in a container cannot manipulate the host because the processes in the container run in an environment isolated from the host.
The isolation is provided by Linux kernel features such as namespaces\cite{man-namespaces} and capabilities\cite{man-capabilities}.
Therefore, a process in a container cannot perform operations including malicious operations on the host side without explicit configurations.
However, vulnerabilities in the low-level\cite{CVE-2019-5736}\cite{runc-1962} or high-level\cite{cve-2019-14271}\cite{containerd-vulrn} runtimes may allow processes in a container to manipulate the host side\cite{container-sec}.
Since the container is running with root privileges, 
the vulnerability in the container runtime can be exploited to run a malicious code with root privileges on the host side.

Rootless containers\cite{rootless-containers} improves the security of containers by running container runtimes without root privileges.
Even if the runtime has vulnerabilities, malicious operations or malicious codes cannot be executed with root privileges.
Rootless containers use some approaches such as user namespaces to enable container runtimes to run without root privileges.
However, networking components designed for root-privileged runtimes cannot be used in rootless containers.
Therefore, RootlessKit\cite{rootlesskit} and slirp4netns\cite{slirp4netns} provide the same functionality for rootless containers.
These components cause huge networking overhead and result in low communication throughput in rootless containers.

\begin{table*}[t!]
\centering
\caption{Comparison with socket replacing related works}
\label{tbl:comparison-previous-works}
\begin{tabular}{c|c|c|c|c|c}
& \begin{tabular}[c]{@{}c@{}}Fully\\ Rootless\end{tabular} 
& Performance
& \begin{tabular}[c]{@{}c@{}}No kernel\\ modification\end{tabular} 
& \begin{tabular}[c]{@{}c@{}}No application\\ modification\end{tabular} 
& \begin{tabular}[c]{@{}c@{}}Publicly\\ available\end{tabular} \\ \hline \hline
\begin{tabular}[c]{@{}c@{}}RootlessKit's port driver\cite{rootlesskit}\\slirp4netns\cite{slirp4netns}\end{tabular} & \checkmark & Low & \checkmark  & \checkmark & \checkmark \\ \hline
lxc-user-nic\cite{lxc-user-nic} & & Mid & \checkmark & \checkmark & \checkmark \\ \hline
AF\_GRAFT\cite{10.1145/3230718.3230723} &  & High &  & LD\_PRELOAD & \checkmark \\ \hline
Slim\cite{227669} & & High & & LD\_PRELOAD & \checkmark \\ \hline
O2H\cite{9860379} & & High & \checkmark & \checkmark & \\ \hline
\textbf{bypass4netns} & \checkmark & High & \checkmark & \checkmark & \checkmark \\ \hline
\end{tabular}
\end{table*}

In this paper, we propose \textbf{bypass4netns}, a method to bypass the communication bottleneck in rootless containers. Theoretically, bypass4netns provides rootless networking with equivalent or better performance to that of a rootful container.
bypass4netns proposes \textbf{socket switching} to bypass the bottleneck for the communications from inside a container to the outside and from the outside to the inside of the container.
It switches a socket created in the process in the container to a socket created on the host.
Socket switching is sometimes called "\emph{socket grafting}"\cite{10.1145/3230718.3230723} and "\emph{socket replacement}"\cite{9860379}.
These methods employed Linux kernel module or eBPF, but it requires root privileges.
bypass4netns employes \textsf{Seccomp User-space Notification~(Seccomp Notify)}\cite{man-seccomp-notify} to switch sockets without application and kernel modifications.
If socket switching is needed, bypass4netns switches sockets with \texttt{SECCOMP\_IOCTL\_NOTIF\_ADDFD} without root privileges.

Containers are often used in a container orchestration system link Kubernetes\cite{kubernetes}.
Usernetes\cite{usernetes} provides the same system with rootless containers, but the communication performance is very low.
This is because it uses VXLAN to provide communication between nodes~(multi-node communication).
Rootless networking components process the VXLAN packets, and it causes huge performance degradation.
To solve this problem, we propose high-performance multi-node communication with bypass4netns.
It does not use any packet encapsulation approaches including VXLAN.
Of course, the method does not require root privilege.

To enable accurate socket switching, we need to handle the syscalls related to the sockets.
Previous works\cite{10.1145/3230718.3230723}\cite{9860379}\cite{227669}, which accelerate container networks by modifying sockets, did not investigate the behavior of the sockets.
They handle syscalls or sockets ad-hoc, but such approaches can cause compatibility issues.
We investigated socket behavior in real applications and implemented bypass4netns based on the traced socket behavior.

We confirmed that bypass4netns can provide 30x faster throughput than that of rootless containers without bypass4netns.
In addition, we evaluated performance with applications and it showed large improvements on some applications.
In summary, our main contributions are that
\begin{enumerate}
    \item We propose an approach to analyze socket behavior to provide highly compatible socket switching.
    \item We propose bypass4netns to accelerate rootless container networking with socket switching. 
    \item We propose a high-performance \textbf{rootless} multi-node communication method.
    \item We implemented bypass4netns including integration \\ with nerdctl\cite{nerdctl}. We confirmed that bypass4netns and the multi-node communication method can accelerate applications' communications without modifying them. 
    As for throughput, bypass4netns provides more than 30x faster networking.
\end{enumerate}

\section{Backgrounds}\label{sec-background}

\subsection{Rootless Containers}
Rootless containers improve the container's security by running its runtime without root privileges.
If the rootless container runtime has vulnerabilities that allow malicious codes to be executed outside of containers, they are executed without root privileges.
The damage will be less than that executed in container runtime with root privileges.

Rootless containers uses \textsf{user namespaces~(UserNS)} to create namespaces without root privileges.
UserNS maps a non-root user like \texttt{uid=1000} in a host to a fake root user (\texttt{uid=0}) in an UserNS.
In the host environment, the user is mapped to a non-root user.
UserNS enables container runtimes to provide containers without root privileges.
However, operations related to network interfaces cannot be achieved with UserNS alone.
This is because that container runtime needs to create a veth pair both in a container and a host.
This veth pair is the entrance and exit point for container communication and requires root privileges to create in a host.
To achieve container networking without root privileges, rootless containers have networking components called \textsf{RootlessKit} and \textsf{slirp4netns}.

\begin{figure}[!t]
  \centering
  \includegraphics[width=0.7\hsize]{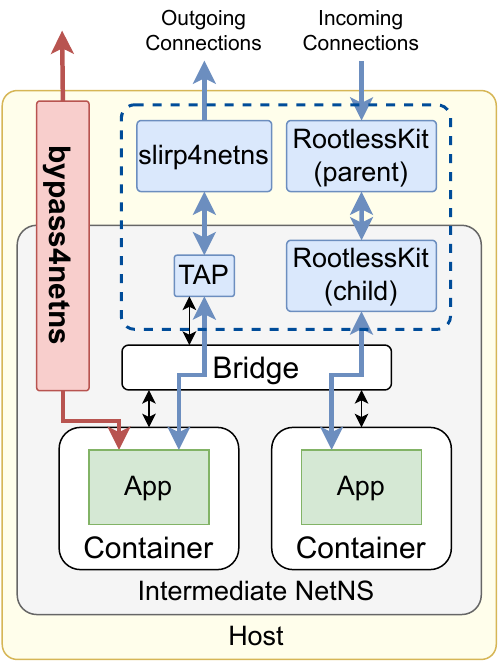}
  \caption{The overview of rootless container network\\Components in blue box provides rootless networking}
  \label{fig:overview-of-rootless-network}
\end{figure}
\cref{fig:overview-of-rootless-network} shows the current rootless containers network.
RootlessKit's port driver\cite{rootlesskit} handles connections from external endpoints.
The port driver in a host accepts connections and the port driver in intermediate NetNS connects to containers.
The port drivers relay connections by simply copying messages between two connections.
slirp4netns\cite{slirp4netns} handles connections from containers to external endpoints.
slirp4netns receives packets from containers via the TAP device and extracts payloads from them.
Then, slirp4netns simply creates a \texttt{SOCK\_STREAM} socket and sends the payloads.
When slirp4netns receives payloads, it constructs TCP packets and sends them to containers via the TAP device.

\subsection{Seccomp}
Seccomp~(secure computing mode)\cite{man-seccomp} is a Linux kernel's functionality for controlling syscall executions.
Seccomp is combined with the container runtime to control syscalls that are executed by processes in a container.
Since Linux kernel 5.9, \texttt{SECCOMP\_IOCTL\_NOTIF\_ADDFD} was added.
This allows the target process to use the file descriptor that specifies the supervisor's file description like a socket.

\subsection{Socket Switching}\label{subsec-socket-switching}
\cref{tbl:comparison-previous-works} shows the comparison between current rootless containers network implementations and other socket switching approaches.
lxc-user-nic\cite{lxc-user-nic} is known as a network feature for LXC unprivileged containers\cite{lxc}.
It can be launched by a non-root user to create a veth pair.
However, lxc-user-nic effectively runs as the root user.
So, if lxc-user-nic has a vulnerability, malicious operations or codes can be executed with root privileges.
Actually, lxc-user-nic had some vulnerabilities\cite{cve-2017-5985}\cite{cve-2018-6556}.
Exact rootless containers should not use lxc-user-nic and the same approach in terms of security.
Also, lxc-user-nic's veth pair uses the container's protocol stack.
It causes some performance degradation\cite{10.1145/3230718.3230723}\cite{227669}\cite{9860379} as same as rootful containers.

Some approaches bypass container networks to accelerate TCP/IP communications.
AF\_GRAFT\cite{10.1145/3230718.3230723} and Slim\cite{227669} replace sockets with their sockets with LD\_PRELOAD.
The shared library specified in LD\_PRELOAD is loaded at first, and the library can hook functions provided by other shared libraries.
libc is a popular shared library that provides socket-related functions and LD\_PRELOAD approaches hook these libc's socket-related function calls and modify their behavior.
However, statically linked applications' sockets cannot be replaced.
Also, they require their kernel module to be loaded.
If kernel modules have vulnerabilities, it causes serious damage to the entire host.
Dependency on a custom kernel module is far from the philosophy of rootless containers.
O2H\cite{9860379} switches sockets in containers to the host's sockets using eBPF.
It does not require modifications to applications including statically linked ones.
However, operations to attach eBPF programs to the host require root privileges.
In rootless containers, all components including the network plugin itself must run without any root privileges.

\section{Analyzing Socket Behavior With eBPF}\label{sec-analysis}
\begin{figure}[!t]
  \centering
  \includegraphics[width=\hsize]{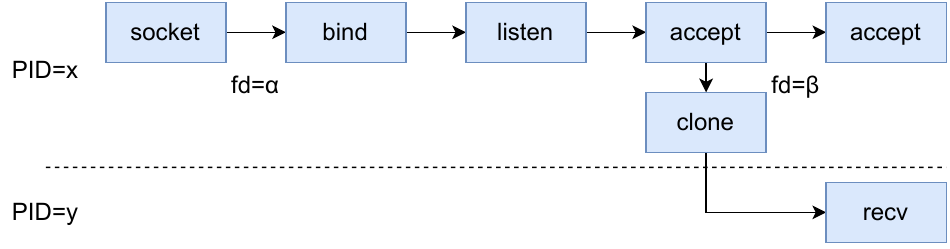}
  \caption{Example of socket tracing with file descriptor}
  \label{fig:socket-trace}
\end{figure}
bypass4netns switch the container's sockets to the host's sockets only if the destination is external endpoints or the binding port is configured to publish.
External endpoints mean endpoints that exist outside of intermediate NetNS.
Simply switching all sockets to the host's sockets increases security concerns.
To choose sockets to switch, bypass4netns needs to know the endpoint of the connection or the port to bind.
Also, some applications configure options to sockets like non-blocking options.
Incorrect socket configurations can make applications behave unexpectedly or cause performance issues.
To ensure high compatibility, bypass4netns need to handle such options precisely and configure sockets.
To implement bypass4netns, we need to investigate the following socket behaviors.

\noindent \textbf{Q1: What kind of syscalls are executed for sockets?}
Handling every syscall degrades application performance.
To minimize performance degradation, bypass4netns need to handle syscalls only related to sockets.
Syscalls perform operations on sockets with file descriptors in their arguments.
Some syscalls(e.g. \texttt{read(2)}) are used not only for sockets.
Therefore, syscalls for sockets are not explicitly given.

\noindent \textbf{Q2: What are the syscalls that configure sockets?}
Applications sometimes configure the non-blocking option or buffer size to their socket.
To avoid affecting application behavior, bypass4netns records options and configures created sockets.
We investigated what syscalls configure options.


We investigated the above points by collecting the syscalls in some applications(iperf3, wget, curl, nginx, apache).
To collect the syscalls, we used \textsf{tracee}\cite{tracee}, which is an eBPF-based syscall tracer.
We implemented the analysis script to retrieve the operations on each socket's file descriptor from the collected syscalls trace.
\cref{fig:socket-trace} shows an example of traced socket behavior.
This script outputs the series of syscall executions on each socket.
The target sockets are \texttt{SOCK\_STREAM}.
This script also takes care of \texttt{fork(2)} and \texttt{clone(2)} since file descriptors are replicated between processes by \texttt{fork(2)} and \texttt{clone(2)},

\begin{table}[t]
\centering
\caption{Traced syscalls related to socket}
\label{tbl:socket-related-syscalls}
\begin{tabular}{c|c}
& syscalls \\ \hline 
(1) Creation      & socket \\ \hline
\textbf{(2) Configuration} & \textbf{fcntl, setsockopt, ioctl} \\ \hline
\textbf{(3) Connection}    & \textbf{connect, bind} \\ \hline
\textbf{(4) Status}        & \textbf{\begin{tabular}[c]{@{}c@{}}getsockopt, getsockname, \\ getpeername\end{tabular}} \\ \hline
(5) Derivation         & accept, accept4, clone \\ \hline
\begin{tabular}[c]{@{}c@{}}(6)\\Communication\end{tabular} & \begin{tabular}[c]{@{}c@{}}poll, recvfrom, sendfile\\ write, select, read, listen \\ lseek, readv, writev, \\ epoll\_ctl, epoll\_wait\end{tabular} \\ \hline
\textbf{(7) Close}         & \textbf{close, shutdown} \\ \hline
\end{tabular}
\end{table}
We investigated the syscalls and classified syscalls into the following 7 classes based on their behavior as shown in \cref{tbl:socket-related-syscalls},
\textbf{(1)~Creation},
\textbf{(2)~Configuration},
\textbf{(3)~Connection},
\textbf{(4)~Status}.
\textbf{(5)~Derivation}
\textbf{(6)~Communication},
\textbf{(7)~Close}.
bypass4netns does not hook (1)~Creation syscalls because bypass4netns cannot know what file descriptor is assigned to the sockets.
The detail is discussed in \cref{subsec-impl-pitfalls}.
As for (2)~Configuration syscalls, the application configures the socket and bypass4netns records the configured options.
The application specifies the destination address~(endpoint) in (3)~Connection syscall's arguments. 
According to the endpoint and container network environment, bypass4netns decides whether to switch the socket or not.
(4)~Status syscalls are used to get socket status like peer address.
bypass4netns handles them and returns dummy values when needed.
(5)~Derivation syscalls are creating new file descriptors from existing sockets or replicating file descriptors into other processes or threads.
As same as (1)~Creation syscalls, bypass4netns does not handle these syscalls directly.
When a new file descriptor appears, bypass4netns checks file descriptors denote newly created sockets or derived from existing sockets.
(6)~Communication syscalls are the core of data transfer and overheads in them affect its communication performance.
Handling syscalls with seccomp causes overheads.
To avoid performance degradation, bypass4netns does not handle (6)~Communication syscalls.
Some previous works\cite{10.1145/3230718.3230723}\cite{227669} handle them to use their own socket APIs.
bypass4netns switches sockets without changing file descriptors and socket APIs~(syscalls), so there is no need to hook (6)~Communication syscalls.
bypass4netns hooks (7)~Close syscalls and cleanup sockets and recorded options.

\section{The Design of bypass4netns}\label{sec-design}
We propose bypass4netns, a method to improve communication performance by switching sockets.
RootlessKit and slirp4netns relay communications between the host and an intermediate NetNS with a daemon or tap device.
bypass4netns bypass an intermediate NetNS by switching sockets created in the container to sockets created on the host side, without using a tap device.

\subsection{The Overview of bypass4netns}
\begin{figure}[t]
    \centering
    \includegraphics[width=\hsize]{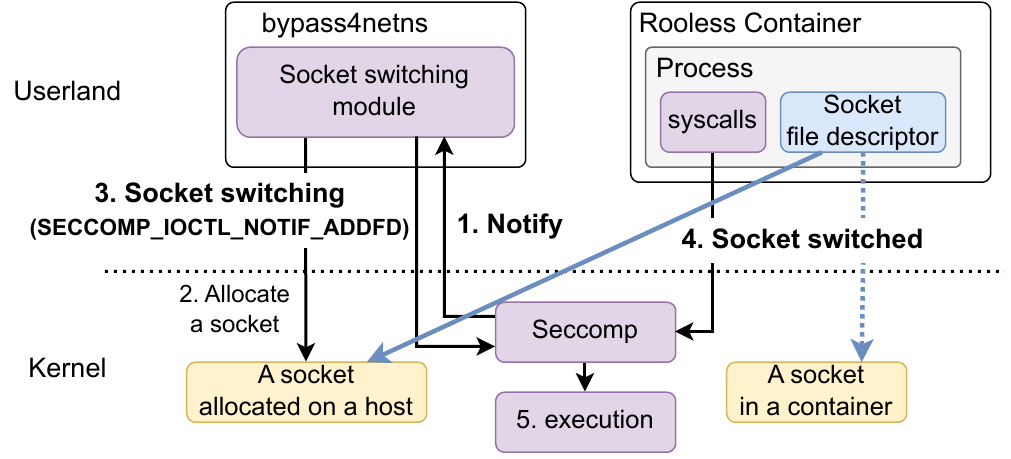}
    \caption{The overview of bypass4netns}
    \label{fig:overview-of-bypass4netns}
  \end{figure}
Fig.~\ref{fig:overview-of-bypass4netns} shows the overview of bypass4netns.
bypass4netns works with the rootless container runtime and uses seccomp user notification~(Seccomp Notify) for socket switching.
The two major roles of bypass4netns are \textbf{the creation of sockets on the host} and \textbf{the switching container's sockets to the host's sockets}.
bypass4netns waits for seccomp notifications and processes them depending on the kind of syscalls.
bypass4netns handles only communications to external endpoints.
Inter-container communication is not handled by bypass4netns because it uses veth pairs and bridges in the intermediate NetNS except communications with switched sockets.
Therefore, bypass4netns handles (3)~Connection syscalls in \cref{sec-analysis} to decide whether to switch a socket or not.
If the communication is to the external endpoint, bypass4netns performs the socket switching.

bypass4netns uses sockets created on the host, and this behavior is similar to an option \texttt{--net=host}.
An option \texttt{--net=host} means that a container uses a NetNS as same as the host.
However, using this option results in leaking abstract unix domain sockets from the host network namespace into containers\cite{do-not-use-net-host}.
To ensure security, bypass4netns does not use the host's NetNS directly and switches sockets selectively.
This design also allows assigning unique IP addresses to each of the containers for inter-container communications.

\subsection{Socket Switching With Seccomp}
As shown in \cref{fig:overview-of-bypass4netns}, bypass4netns performs socket switching with \texttt{SECCOMP\_IOCTL\_NOTIF\_ADDFD}.
bypass4netns receives a notification message containing information about the syscall via notification fd.
After the validation, bypass4netns reads the information about the syscall.
For syscalls that are not related to socket switching, a response message that allows syscall execution is sent.
If a syscall is related to the socket, bypass4netns processes it.

\begin{figure}[t!]
    \centering
    \includegraphics[width=0.7\hsize]{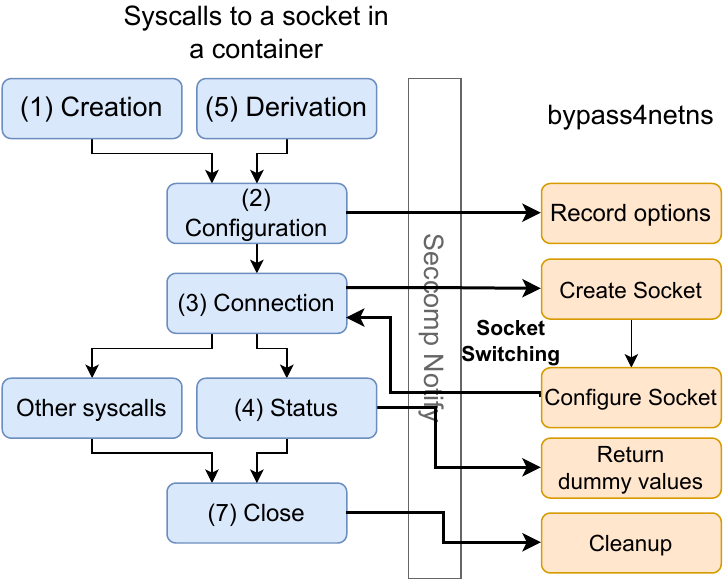}
    \caption{Flow of socket switching}
    \label{fig:flow-bypass4netns}
\end{figure}
\cref{fig:flow-bypass4netns} shows the flow of socket switching.
bypass4netns reads (2)~Configuration syscalls' arguments and records them as socket's options.
When (3)~Connection syscalls are executed, bypass4netns reads their destination address or biding port and switches sockets if the destination is an external endpoint or the port is published.
In socket switching, bypass4netns creates a new socket.
bypass4netns configures the new socket based on the record to make the socket's state to same as the container's socket.
Then, using \texttt{SECCOMP\_} \texttt{IOCTL\_NOTIF\_ADDFD},
bypass4netns switches the container's socket to the created socket.

Once the socket is switched, other containers cannot connect to the application.
This is because the socket is bound to the published port.
\cref{fig:flow-bypass4netns} describe the situation.
Container A's application has the socket~(80/tcp) and the socket is published as \texttt{8080/tcp}.
Container B tries to connect to the application with an address \texttt{ContainerA:8080/tcp}, but it fails because the socket is already switched and bound to \texttt{80/tcp}.
bypass4netns is designed to handle such container-to-container communication.
When container B tries to connect to the switched socket, bypass4netns \textbf{rewrites the destination address}.
bypass4netns reads the address and rewrite the destination address \texttt{ContainerA:80/tcp} to 

\noindent \texttt{Host:8080/tcp} by writing \texttt{/proc/<pid>/mem}.
Then, the process continues to connect to the switched socket and it will succeed.
After the socket switching, the process sometimes calls \texttt{getpeername(2)}.
bypass4netns handle it and return \texttt{ContainerA:80} as the peer.

\subsection{Policy-aware Socket Switching}\label{subsec-probe}
\begin{figure}[t]
  \centering
  \includegraphics[width=\hsize]{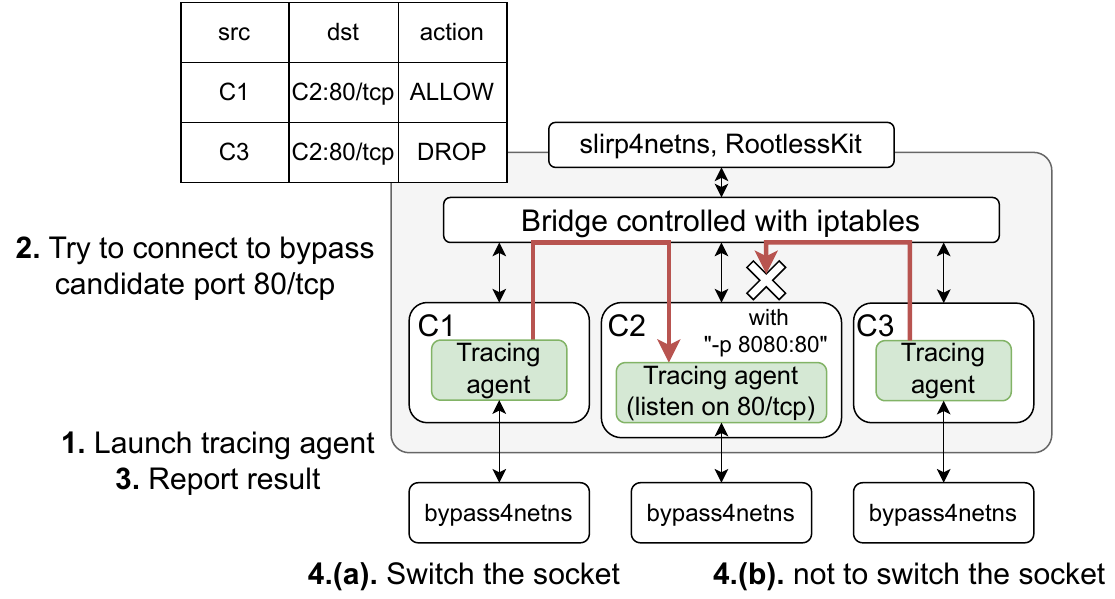}
  \caption{Switching decision with probe agents}
  \label{fig:tracer-agent}
\end{figure}
Users sometimes configure container networks to prohibit specific container communication.
Also, users can create multiple container networks and configure a routing table to allow containers to communicate between the networks.
To apply socket switching securely, we need to know whether the container can communicate with other containers or not.
bypass4netns needs to decide whether the socket can be switched or not.
If a wrong decision is made, it unexpectedly allows communication between containers.

To collect connectivity between containers, a static approach and a dynamic approach can be used.
A static approach judges the communication is allowed or not with iptables policy and network interface information including routing table.
This approach does not require active probes and does not affect the container network environment.
However, if other components are not considered in this approach to control the container network, the decision will be wrong.
Also, such a decision mechanism will become complicated.

To make correct decisions with a simple mechanism, we employed a dynamic approach.
\cref{fig:tracer-agent} describes the dynamic approach with probe agents.
A dynamic approach uses a probe agent in each container and checks the connectivity actively.
Probe agents are launched in the same NetNS as each container and communicate probe results with bypass4netns.
Each probe agent binds sockets to published ports and tries to connect to other container's published ports.
If the socket connects to other probe agents successfully, the sockets between these containers can be switched.
If the socket fails to connect, the sockets are not switched.
To reduce socket switching latency, the probes check the connectivity periodically and cache the results.

\subsection{Accelerating Multi-node Communication}\label{subsec-multi-node}
\begin{figure}[t]
  \centering
  \includegraphics[width=0.8\hsize]{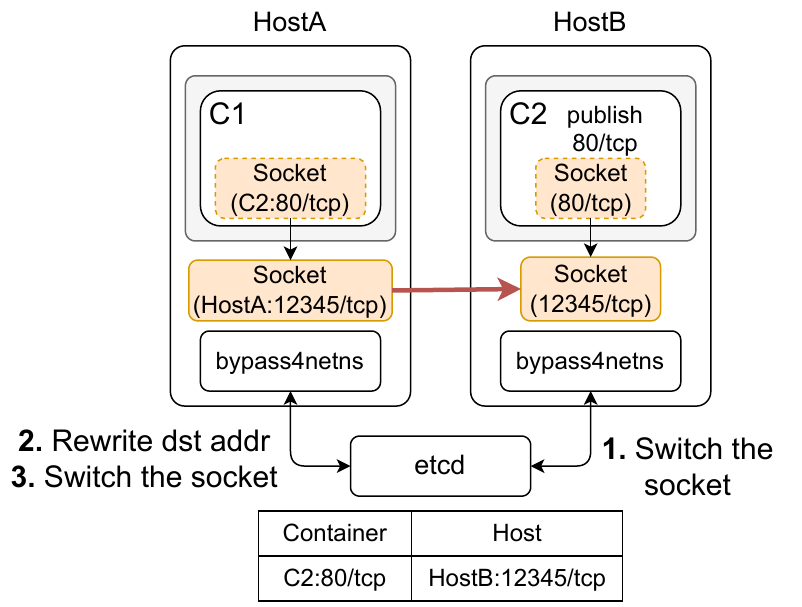}
  \caption{Multi-node communication with bypass4netns}
  \label{fig:multinode}
\end{figure}
Containers are often used not only in a single-node environment but also in a multi-node environment like Kubernetes.
As for multi-node communication, users often use overlay networks with encapsulation approaches\cite{flannel} or L3 routing-based approaches\cite{calico}.
Usernetes\cite{usernetes}, which is a Kubernetes in rootless containers, provides multi-node communication with VXLAN.
In Usernetes, the VXLAN interface is created in rootless containers to create the interface without root privilege.
The interface allocates sockets in the kernel, but bypass4netns handles only user processes' sockets and cannot handle the interface's socket. 
VXLAN packets are handles are handled by slirp4netns and RootlessKit.
Due to the overhead in them, the performance between nodes becomes lower than port-forwarded communication and rootful container's VXLAN communication.

To provide faster multi-node communication for rootless containers, bypass4netns provides a simple multi-node communication mechanism without VXLAN as same as Slim.
\cref{fig:multinode} describes the mechanism.
Each node has rootless containers and some ports are published.
Each bypass4netns shares the relation between the published container's address and the host's addresses via KVS. 
We use \textsf{etcd} as a KVS.
When container C1 on HostA tries to connect to the published port of container C2 on HostB, bypass4netns rewrites its destination address to the corresponding HostB's address.

\section{Implementation}\label{sec-implementation}
We implemented bypass4netns based on the analysis in \cref{sec-analysis} and the design in \cref{sec-design}.
We also implemented \textsf{bypass4netnsd} and some patches in \textsf{nerdctl} to manage bypass4netns.
The implementation is available at \url{https://github.com/rootless-containers/bypass4netns}.

\noindent\textbf{bypass4netns }
bypass4netns receives notifications from containers and switches sockets as needed.
\\bypass4netns itself is assigned on a per-container basis.

\noindent\textbf{bypass4netnsd }
bypass4netnsd manages bypass4netns' lifecycles and configs.
When a container with bypass4netns starts, bypass4netnsd receives the container and network information from nerdctl such as the port to be published.

\noindent\textbf{nerdctl }
nerdctl\cite{nerdctl} is a Docker-compatible CLI for containerd\cite{containerd}.
It cooperates with containerd and provides container functionality.
In rootless containers, nerdctl configures rootless components including RootlessKit and slirp4netns when a container starts or stops.
When a container with bypass4netns starts, nerdctl notifies information about the container to bypass4netnsd and waits for bypass4netns to start.
Also, when the container stops, nerdctl notifies bypass4netnsd and waits for bypass4netns to stop.

\subsection{Insights in Implementation}\label{subsec-impl-pitfalls}
\noindent\textbf{Seccomp Notify can handle only syscalls before execution}

Seccomp is a module to filter syscalls before execution.
It is also true for Seccomp Notify.
It checks syscalls and decides whether to execute or not dynamically before execution.
bypass4netns manages options and states depending on file descriptors.
However, bypass4netns cannot know what a file descriptor is assigned to the socket because bypass4netns cannot handle \texttt{socket(2)} return value due to Seccomp's limitation mentioned above.
To overcome this limitation, bypass4netns checks whether the file descriptor is a socket or not and registers the file descriptor when syscalls are hooked.
In the implementation, bypass4netns gets the file descriptor with \texttt{pidfd\_getfd(2)} and tries to get the socket type with \texttt{getsockopt(2)}.
If it fails or the socket type is not \texttt{SOCK\_STREAM}, the file descriptor is registered as a non-switchable socket.
Also, file descriptors sometimes denote already connected sockets.
Such file descriptors can be seen after (5)~Derivation syscalls in \cref{fig:flow-bypass4netns}.
bypass4netns checks whether the socket is connected or not with \texttt{getpeername(2)}.
If the socket is connected, register the corresponding file descriptor as a non-switchable socket.
Once the file descriptor is registered, bypass4netns performs socket switching based on the registered state and recorded options.


\noindent\textbf{Insufficient privileges}

bypass4netns uses \texttt{/proc/<PID>/mem} to read and write destination address.
Some container applications create their users in a container and run with the user.
bypass4netns runs with a normal user's privilege and does not have the permissions to read or write the other user process's memory.
To read and write other process's memory, bypass4netns launches an agent to open \texttt{/proc/<PID>/mem} with the process's user using \texttt{nsenter(1)}.
The agent passes the opened file descriptor to bypass4netns via Unix Domain Socket, and bypass4netns reads and writes the process's memory.

\begin{figure*}[t]
    \centering
    \begin{minipage}[b]{0.45\hsize}
        \centering
        \includegraphics[width=\hsize]{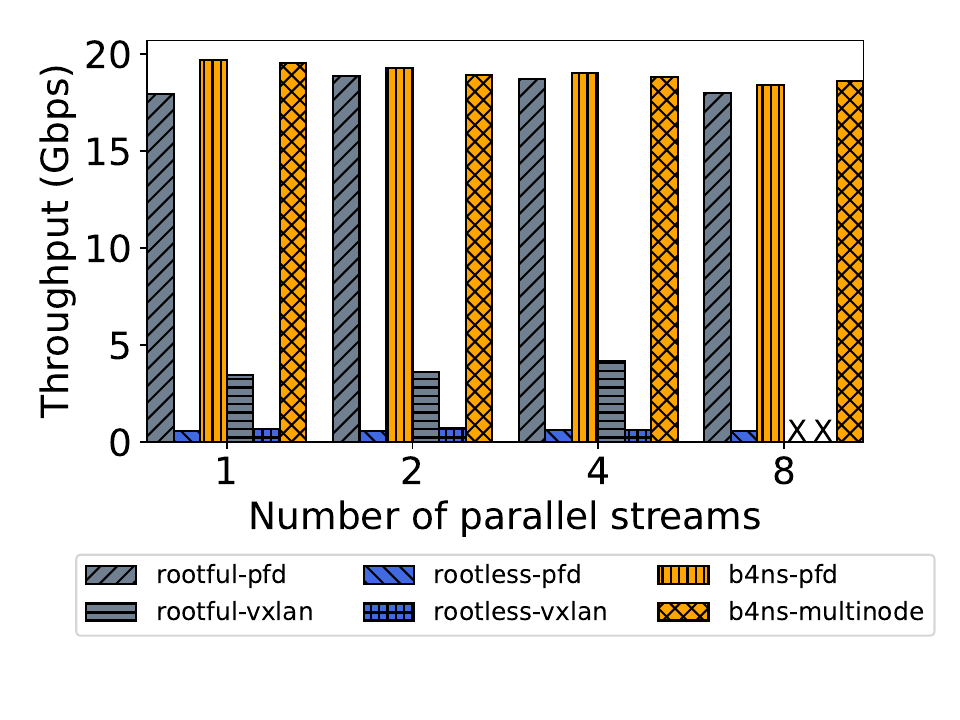}
        \vspace{-1cm}
        \caption{Benchmark result with iperf3}
        \label{fig:eval-iperf3}
    \end{minipage}
    \begin{minipage}[b]{0.45\hsize}
        \centering
        \includegraphics[width=\hsize]{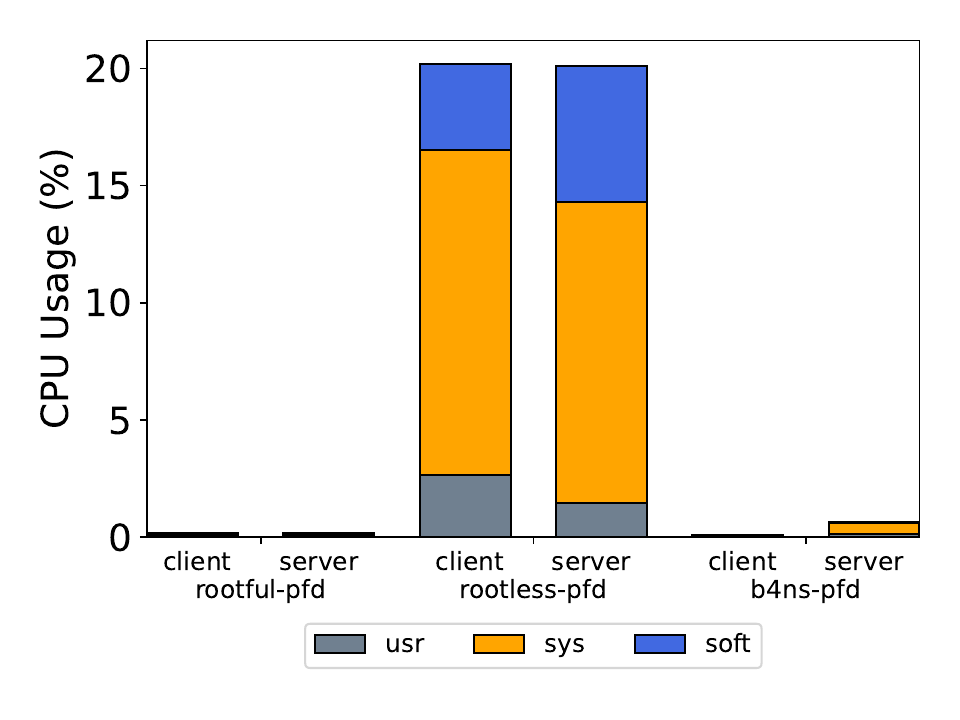}
        \vspace{-10mm}
        \caption{CPU utilization while running iperf3 client}
        \label{fig:performance-cpu-utilization}
    \end{minipage}
\end{figure*}
\section{Evaluation}\label{sec-evaluation}
We evaluated the communication performance in terms of throughput and real application performance.
Evaluation scripts are available at \url{https://github.com/naoki9911/bypass4netns/tree/benchmark}.

\subsection{Environment}
The evaluations are performed on QEMU/KVM virtual machines (VMs).
The host machine has AMD EPYC 7452~(32 Cores, 2.35 GHz base clock) with Ubuntu 22.04.
Each VM is assigned 8 vCPU cores and 16GB memory with Ubuntu 22.04.
The evaluation was performed between 2 VMs.

Performance benchmarks are conducted with 6 networks.
\noindent\textbf{rootful-pfd }
Containers run with containerd running with root.
Ports are published with port forwarding. (i.e. container runs with \texttt{-p} option.)

\noindent\textbf{rootless-pfd }
Containers are rootless containers.
Ports are published with port forwarding, and all communications are processed by slirp4netns and RootlessKit.

\noindent\textbf{b4ns-pfd }
Containers are rootless containers and bypass4netns performs socket switching.
Ports are published with socket switching.

\noindent\textbf{rootful-vxlan }
Containers run with containerd running with root.
Hosts have a VXLAN interface and containers in both hosts can communicate via VXLAN tunnel.

\noindent\textbf{rootless-vxlan }
Containers are rootless containers.
Containers have a VXLAN interface and port \texttt{4789/udp} is published.
It means slirp4netns and RootlessKit process all VXLAN packets.

\noindent\textbf{b4ns-multinode }
Containers are rootless containers and bypass4netns performs socket switching.
bypass4netns runs with multi-node communication enabled.

We performed performance benchmarks for iperf3, nginx, redis, mysql, and rabbitmq with benchmark tools.

\subsection{Application Compatibility}\label{subsec-app-compatibility}
\begin{table}[t]
\centering
\caption{Tested Applications}
\label{tbl:tested-applications}
\begin{tabular}{c|c}
Application (version)         & Language        \\ \hline \hline
iperf3 (3.16)              & C                  \\ \hline
nginx (1.18)               & C                  \\ \hline
redis (7.2.3)                & C                  \\ \hline
memcached (1.6.22)            & C                  \\ \hline
rabbitmq (3.12.10)             & Erlang             \\ \hline
etcd (3.3.25)                 & Go                 \\ \hline
postgresql (16.1)           & C                  \\ \hline
mysql (8.2.0)                & C++                \\ \hline
HTTP Client (Go 1.21.3)          & Go~(static linked)        \\ \hline
redis-benchmark (7.2.3)      & C                    \\ \hline
memtier\_benchmark (2.0.0)   & C++                  \\ \hline
rabbitmq-perf-tester (2.20.0) & Java                 \\ \hline
etcd-benchmark (3.3.25)       & Go~(static linked)        \\ \hline
pgbench (16.1)              & C               \\ \hline
sysbench (1.0.20)             & C           \\ \hline
\end{tabular}
\end{table}
We have tested famous applications.
\cref{tbl:tested-applications} describes tested applications.
To probe bypass4netns works without modifying applications, we used official container images if the developers distributed them.
If not distributed, we built container images from the source code without modifying it.
bypass4netns worked with most of these applications including static linked applications.
However, running \textbf{HTTP Client~(Go 1.21.3)} with more than 2 HTTP Clients with gorutines caused a strange issue.
The clients run concurrently with goroutine and they independently allocated sockets.
In theory, bypass4netns hooks syscalls and performs socket switching, but Seccomp Notify dropped some syscalls silently.
bypass4netns could not perform socket switching correctly and HTTP clients could not communicate with the server.
Seccomp Notify seems to have a thread-related bug and we need to fix the issue to apply bypass4netns in a real environment.
Running only 1 HTTP client did not cause the issue, and we have performed the performance evaluation with 1 HTTP client.

\subsection{Application Performance}
\begin{figure}[t]
    \centering
    \includegraphics[width=\hsize]{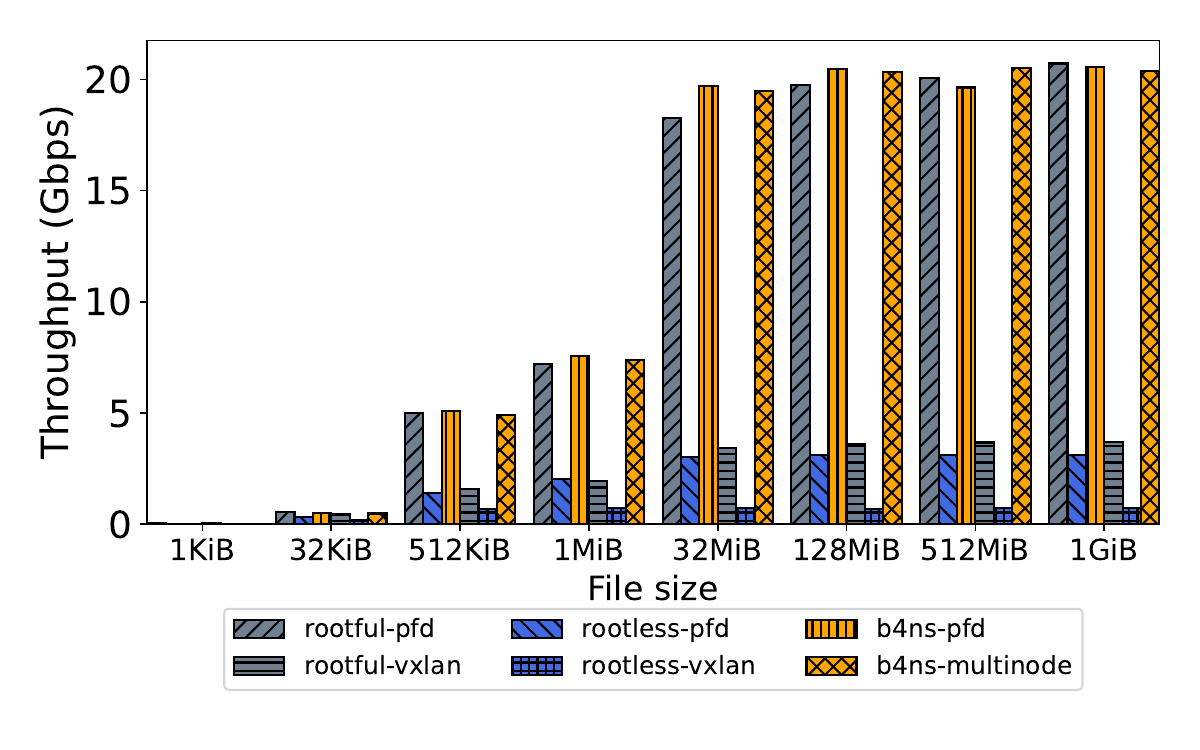}
    \vspace{-1cm}
    \caption{Benchmark result with block download}
    \label{fig:eval-block}
\end{figure}
\begin{figure*}[t]
    \centering
    \includegraphics[width=\hsize]{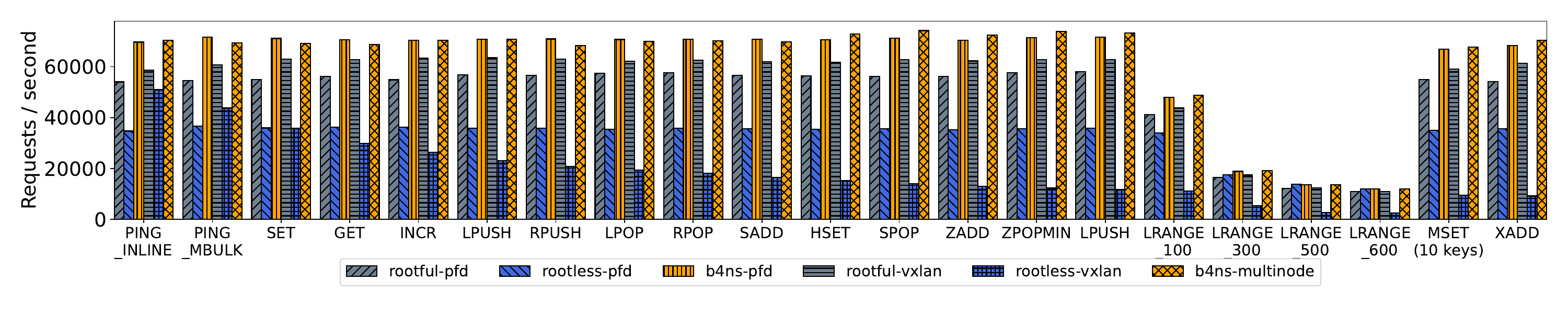}
    \vspace{-0.8cm}
    \caption{Benchmark result with redis-benchmark (Requests per second)}
    \label{fig:eval-redis}
\end{figure*}
\begin{figure*}[t]
    \centering
    \includegraphics[width=\hsize]{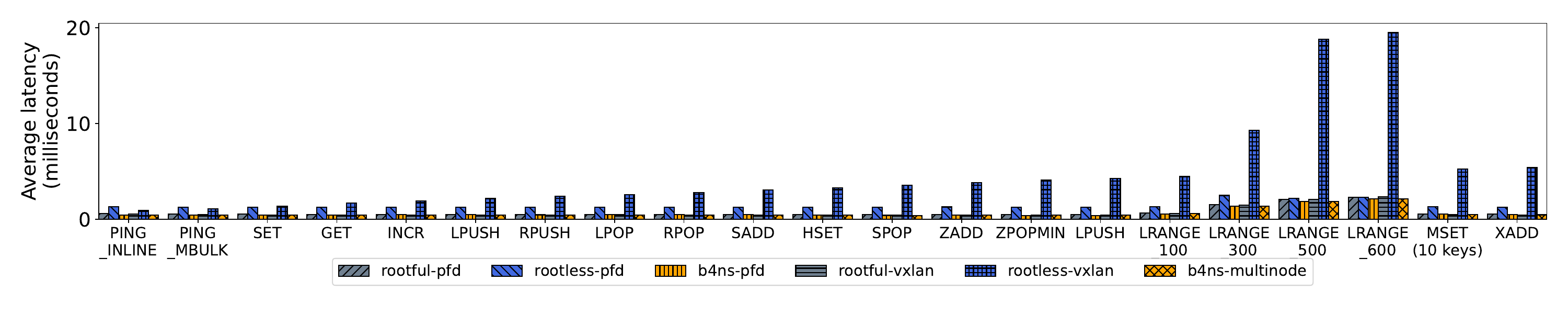}
    \vspace{-0.8cm}
    \caption{Benchmark result with redis-benchmark (Average latency)}
    \label{fig:eval-redis-latency}
\end{figure*}
\begin{figure}[t]
    \centering
    \includegraphics[width=\hsize]{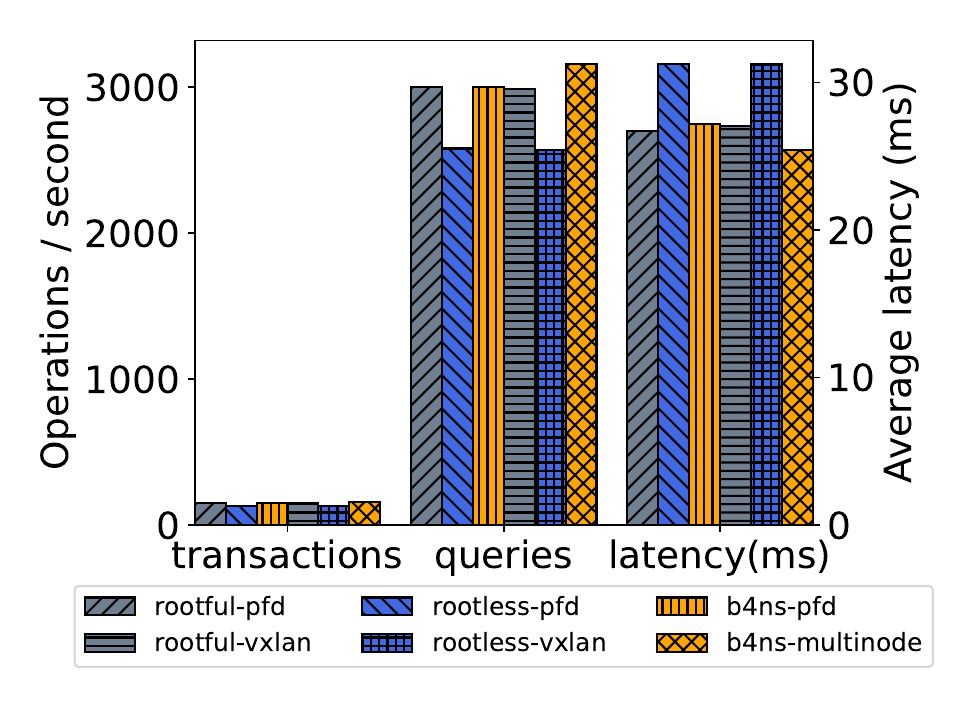}
    \vspace{-1cm}
    \caption{MySQL benchmark result with sysbench}
    \label{fig:eval-mysql}
\end{figure}
\subsubsection{iperf3}
\cref{fig:eval-iperf3} shows the throughput measured by running iperf3 for 120 seconds and the values are average of 10 times measurement.
The other VM runs the iperf3 server.
As for the result with 1 parallel stream,
rootful-pfd~(17.9 Gbps), b4ns-pfd~(19.7 Gbps), and b4ns-multinode~(19.5 Gbps) achieved almost 20 Gbps.
Rootless containers without bypass4netns showed low performance.
rootless-pfd was 570 Mbps and, it did not reach 1Gbps.
A rootless container with bypass4netns provides more than 30 times faster throughput than that without bypass4netns.
This is because slirp4netns, which handles outgoing connections, has a huge overhead to handle packets and convert them as unprivileged syscalls.

We increased iperf3's parallel streams to measure the performance when concurrent TCP connections increased.
From iperf version 3.16, it supports multi-threaded TCP stream processing and actually, it used multiple CPU cores in the evaluation.
The entire result shows the almost same result as in the result with 1 parallel stream.
The evaluation environment was built with 2 VMs and it seems that the performance was limited due to the host's side limitation.
As for rootful-vxlan and rootless-vxlan, iperf3 clients with 8 parallel streams did not finish successfully.
iperf3 has a control channel to manage streams and benchmark results, but the channel broke in these cases.
We consider that overloaded VXLAN encapsulation and decapsulation caused the problem.
Retransmissions per stream in 10 seconds exceeded 1000 times in these cases, and it is very high compared to less than 100 times in 120 seconds on rooful or bypass4netns cases.
This lossy network may have caused the control channel disconnection.
For reference, the throughput during measurement was about 4.47 Gbps for rootful-vxlan and 567 Mbps for rootless-vxlan.
bypass4netns is mainly aimed at improving the performance of TCP/IP communications, and we confirmed that bypass4netns improves their performance.

We measured CPU utilization while running iperf3 with the rate limited to 1Gbps with 1 stream.
The measurement results are shown in Fig.~\ref{fig:performance-cpu-utilization}.
\texttt{usr} means user process CPU usage.
\texttt{sys} means kernel process CPU usage.
\texttt{soft} means software interruption.
CPU utilization was almost 20\% without bypass4netns.
This is due to packet processing in slirp4netns.
On the other hand, when bypass4netns is enabled, the CPU utilization is low.
This is because bypass4netns only switches sockets and the communication itself is handled by the kernel as is in the rootful containers.

The throughput with bypass4netns was slightly faster than that in a rootful container.
Rootful containers configure iptables and CNI plugins for container networking.
bypass4netns bypasses such networking modules and reduces the overhead.
As a result, rootless containers with bypass4netns provide better performance than rootful containers.

As for multi-node communications, The performance for VXLAN-based overlay networks in both rootful and rootless containers was lower than that of port forwarding.
rootful-vxlan was 3.4 Gbps and rootless-vxlan was 658 Mbps with 1 parallel stream.
As previous works\cite{227669}\cite{9860379} mentioned, this is considered to be caused by the VXLAN encapsulation and decapsulation costs and insufficient hardware offloading.
The performance for multi-node communication with bypass4netns was as same as b4ns-pfd.
In multi-node communication, bypass4netns performs socket switching with the same approach when ports are published.
The performance in both b4ns-pfd and b4ns-multinode is theoretically the same performance and the result shows that.

\subsubsection{static file distribution with nginx}
\cref{fig:eval-block} shows the static file distribution performance with nginx.
We measured the time to download files 1000 times and calculated the throughput.
We used files generated by \texttt{/dev/urandom} and placed them in nginx's public directory.
We implemented a benchmark client with Go language and \texttt{net/http} as an HTTP client.
The client is built as a statically linked binary.

The result shows the same tendency in the iperf3 benchmark.
The benchmark worked and it means that the client is statically linked and bypass4netns can handle binaries that LD\_PRELOAD cannot handle.
The performance for rootful-pfd and b4ns-pfd is almost the same.
rootless-pfd, rootless-vxlan, and rootful-vxlan showed low performance and we consider this was caused by the same reason in the iperf3 benchmark.

\subsubsection{redis}

We measured redis performance with redis-benchmark.
Redis is a famous Key-Value Store~(KVS) and it provides a fast database for KV.
We run redis-benchmark with default parameters~(data size is 3 bytes, 50 parallel connections, no multithreads).
\cref{fig:eval-redis} shows the results.
The performance with bypass4netns is almost twice as compared to rootless containers.
\cref{fig:eval-redis} shows average latencies for each operation.
Rootless containers without bypass4netns~(rootless-pfd, rootless-vxlan) had higher latencies than rootful and rootless containers with bypass4netns~(b4ns-pfd, b4ns-multinode).
Like iperf3 and nginx, rootless networking components cause overhead, resulting in higher latencies and lower performance.
Rootful containers~(rootful-pfd, rootful-vxlan) showed lower performance than rootless containers with bypass4netns.
\cref{fig:eval-redis} shows rootful containers had slightly higher latencies.
In benchmarks of iperf3 and nginx, the communications were bulk transfer and it is not affected heavily by latency.
Redis is a simple KVS with request and response-based protocol, and the latency seems to affect its performance drastically.
Networking components like iptables and Linux bridge increase the latency a bit.
Even if the added latency is a little, the performance is heavily affected.
We consider that bypass4netns bypasses such components and that is why redis-benchmark with bypass4netns achieves the best performance.

\subsubsection{mysql}

We measured MySQL performance with sysbench.
MySQL is a famous relational database~(RDB) and sysbench is a multi-threaded benchmark tool for database systems.
We run sysbench for 120 seconds with the following parameters, and the values are 10 times the average.
The number of threads is 4, the running time is 120 seconds, and the test is \texttt{oltp\_read\_write}.
\texttt{oltp\_read\_write} test is a online transaction processing~(OLTP) test with read and write.

\cref{fig:eval-mysql} shows the result.
Rootless containers~(rootless-pfd, rootless-vxlan) performance is relatively lower than others and latency is higher than others.
However, we did not see huge performance improvements as we saw in iperf3, nginx, and redis.
Overall latency is higher compared to the benchmark with redis.
This implies MySQL requires a long time to process transactions or queries on a server.
For such long round-trip time applications, the impact of the improvement in the network was relatively small, and a small improvement was seen in the benchmark.

\section{Discussion}\label{sec-discussion}
\subsection{Compatibility With Existing Applications}
As described in \cref{sec-evaluation}, we showed the performance improvements in some applications.
We also proposed the eBPF-based analysis tool and it will help bypass4netns developers to provide more compatibility.
However, we experienced some strange behavior in Seccomp Notify.
As mentioned in \cref{subsec-app-compatibility}, Seccomp Notify handles all specified syscalls, but it sometimes drops syscalls.
bypass4netns implements some workaround to handle drops of syscalls including re-checking whether the socket is switchable or not.
Our goal is to apply bypass4netns to every application running in rootless containers, but some improvements in Seccomp Notify are required.

\subsection{Integration With Other Components}
bypass4netns is a mechanism to improve communication performance by bypassing intermediate NetNS isolation with sockets created on the host.
If communication policies such as iptables policies are set in the intermediate NetNS, they will be ignored.
Therefore, it is difficult to control or mark communications using iptables or CNI plugins in intermediate NetNS.

\subsection{Kubernetes Support}
We proposed a method to accelerate multi-node communication with bypass4netns in \cref{subsec-multi-node}.
Container orchestration tools like Kubernetes\cite{kubernetes} require plugins to manage their container network.
In Kubernetes network model\cite{kubernetes-network}, it requires 3 communications, Pod-to-Pod, Pod-to-Service, and External-to-Pod.
In this classification, multi-node communication in bypass4netns is Pod-to-Pod communication.
Other communications are handled in the concept of Service.

Service has an IP address called \textsf{Service IP}, providing load balancing for backend Pods with the Service IP.
Currently, Kubernetes implements Service with iptables, but modifying iptables's tables requires root privilege.
Configuring iptables in a rootless container or intermediate NetNS does not require root privilege.
However, it requires a proxy between their NetNS and the host's NetNS and it causes performance degradation.
To provide Service, we need to consider another approach without iptables and a proxy.

One approach is to use bypass4netns as a load balancer.
bypass4netns rewrite \texttt{connect(2)}'s destination in multi-node communication.
Rewriting the destination address~(Service IP) to randomly chosen backend Pods provides pseudo load balancing.

We do not have implemented components for other communications yet, and we also need to implement the dedicated CNI plugin to apply bypass4netns in Kubernetes.
With the dedicated CNI plugin and Service implementation, bypass4netns will provide faster networking in Kubernetes.

\subsection{Security Considerations}
External communications with bypass4netns are treated as same as other communications on the host.
This is because the sockets switched by bypass4netns are created on the host.
Without bypass4netns, as mentioned above, container communications are controlled in the intermediate NetNS.
Instead of access control with iptables, we proposed a dynamic probe approach in \cref{subsec-probe} to reflect their policies in socket switching.
XMasq\cite{lin2023xmasq}, a method to bypass protocol stack, provides a similar approach with eBPF.
Users can control communications indirectly by configuring iptables.
Also, bypass4netns has an interface to control every \texttt{connect(2)}.
This is an original purpose of Seccomp Notify and it provides syscall-based or destination-based access control.

Seccomp Notify has the probability to allow the container to connect to the host's loopback by exploiting a time-of-check to time-of-use(TOCTOU) attack\cite{toctou}.
TOCTOU is the race condition that happens in modifying messages or results after the check but before use.
This may allow a container to connect to other services hosted on the host.
Linux kernel developers are addressing this issue and we continue to watch the situation closely.

\subsection{As a Versatile Socket Switching Approach}
Some faster TCP/IP communication approaches\cite{180259}\cite{179773} provide socket-like API.
To use them, we need to rewrite socket-related function calls in applications.
However, not all applications can be rewritten.
LD\_PRELOAD enables applications to use modules without rewriting.
As we mentioned in \cref{subsec-socket-switching}, static-linked applications cannot use that.
bypass4netns provides functionality to switch sockets without modifying applications and LD\_PRELOAD.
In this paper, we demonstrated its functionality with rootless containers and used Linux's TCP/IP stack as the target sockets, but it can be used for other purposes.
We think that bypass4netns can switch Linux's sockets to other custom module sockets if the module provides BSD-socket API and file descriptor-based sockets.
Also, we proposed the socket tracing method to investigate socket APIs used in real applications and their behaviors.
The methods we proposed in this paper enable us to develop high-compatible TCP/IP stacks and bypass4netns makes it easy to use newly developed stacks with existing applications. 

\section{Conclusion}\label{sec-conclusion}
In this paper, we propose bypass4netns, a method to improve the TCP/IP communication performance to or from the external endpoints in rootless containers.
In a rootless container, the network component to relay communications between the intermediate NetNS for the rootless container and the host is the bottleneck.
bypass4netns switches the socket in the container to the socket created on the host and bypasses the slow networking components.
We implemented bypass4netns and confirmed that it can achieve communication 30x faster than conventional rootless containers without modifications to applications and kernel.
Future works exist including Kubernetes integration and debugging Seccomp Notify's slicent syscall drop.
We will continue to work to provide faster and more compatible networking in rootless containers.

The implementation is available at \url{https://github.com/rootless-containers/bypass4netns}.

\bibliographystyle{abbrv}
\bibliography{main}

\end{document}